\documentclass[preprint,12pt,authoryear]{elsarticle}

\usepackage{amssymb}

\usepackage{amsmath}

\journal{Icarus}

\begin{document}

\begin{frontmatter}

\title{Decoupling Jupiter's deep and atmospheric flows using the upcoming
Juno gravity measurements and a dynamical inverse model}

\author{Eli Galanti, Yohai Kaspi}

\address{Weizmann Institute of Science, Rehovot, Israel}

\begin{abstract}
Observations of the flow on Jupiter exists essentially only for the
cloud-level, which is dominated by strong east-west jet-streams. These
have been suggested to result from dynamics in a superficial thin
weather-layer, or alternatively be a manifestation of deep interior
cylindrical flows. However, it is possible that the observed winds
are indeed superficial, yet there exists deep flow that is completely
decoupled from it. To date, all models linking the wind, via the induced
density anomalies, to the gravity field, to be measured by Juno, consider
only flow that is a projection of the observed could-level wind. Here
we explore the possibility of complex wind dynamics that include both
the shallow weather-layer wind, and a deep flow that is decoupled
from the flow above it. The upper flow is based on the observed cloud-level
flow and is set to decay with depth. The deep flow is constructed
to produce cylindrical structures with variable width and magnitude,
thus allowing for a wide range of possible scenarios for the unknown
deep flow. The combined flow is then related to the density anomalies
and gravitational moments via a dynamical model. An adjoint inverse
model is used for optimizing the parameters controlling the setup
of the deep and surface-bound flows, so that these flows can be reconstructed
given a gravity field. We show that the model can be used for examination
of various scenarios, including cases in which the deep flow is dominating
over the surface wind. We discuss extensively the uncertainties associated
with the model solution. The flexibility of the adjoint method allows
for a wide range of dynamical setups, so that when new observations
and physical understanding will arise, these constraints could be
easily implemented and used to better decipher Jupiter flow dynamics.
\end{abstract}

\end{frontmatter}

\section{Introduction}

The nature of the flow on Jupiter below the observed cloud-level is
still mostly unknown. Analysis of the cloud-level flow, based on tracking
of cloud observations \citep[e.g.,][]{Porco2003}, shows strong east-west
flow of up to 140 m~s$^{-1}$, with some local non zonal flows such
as around the Great Red Spot. Below the cloud-level, the Galileo probe
\citep{Atkinson1996} showed winds of 160 m~s$^{-1}$ going down
to a depth of at least 24~bars at a specific location ($6^{\circ}$N),
but it is questionable of whether this represents the general flow
\citep{Orton1998,Showman2000}. Some studies suggest, based on indirect
observations, that non-zero winds should exist below the cloud-level
\citep{Conrath1981,Gierasch1986,Dowling1988,Dowling1989}, but their
conclusions were limited to a depth of less than 1\% of the planet's
radius.

Theoretical understanding and numerical modeling during the past decades
can be divided into two general mechanistic approaches. The first,
assumes the flow is confined to a shallow region, close to the cloud-level,
similar to atmospheres of terrestrial planets, and becomes organized
into zonal jets due to atmospheric turbulence \citep{Rhines1975,Rhines1979}.
The energy source for the flow can then either come from internal
heating or solar radiation. The mechanism governing such shallow zonal
flows was suggested to be either turbulence forced from the lower
layers \citep[e.g.,][]{Williams1978,Williams2003,Showman2007,Kaspi2007},
or shallow decaying turbulence \citep[e.g.,][]{Cho1996,Scott2007}.
Other studies, using idealized general circulation models solving
for the full primitive equations, were even able to simulate cloud-level
flow structures that are consistent with those observed in all solar
system giant planets \citep{Lian2010,Liu2010}. The second approach
assumes that the observed cloud-level flow is a surface manifestation
of convective columns originating from the hot interiors of the planet
\citep{Busse1976,Busse1994}. Angular momentum conservation in a rapidly
rotating planet like Jupiter leads the flow to be aligned with the
direction of the spin axis, and it has been was shown in many studies
that strong internal convection can lead to zonally symmetric flows
aligned parallel to the axis of rotation \citep[e.g.,][]{Aurnou2001,Christensen2002,Wicht2002,Heimpel2005,Kaspi2009,Jones2009,Gastine2012,Gastine2013,Chan2013}.
In all these studies, however, the width of the equatorial east to
west super-rotation is much greater than that observed on Jupiter.
These two approaches have been in debate for the last 40 years with
no observed data that could resolve the controversy. A third option,
not considered in previous studies, is that both type of flows exist
alongside: an internal flow of an unknown character and likely forced
by convection and shallow flow related to the observed cloud level
winds. Such a scenario would require additional dynamics existing
beneath the cloud level so that the weather-layer winds would decay
with depth (e.g., due to latent heat release, or enhanced stratification
at the radiative-convective boundary), while the deep winds will occupy
the deep convective region which is unaffected by the solar radiation.

The expected gravity measurements of Jupiter by Juno might give additional
information about the character of the flow. Starting in the fall
of 2016, the Juno spacecraft will perform high accuracy gravity measurements,
with sensitivity expected to allow measurements at least up to gravity
harmonic $J_{10}$ \citep{Bolton2005,Finocchiaro2010}. Several studies
have shown that these gravity measurements could be used to decipher
the flow on the planet below its cloud-level \citep{Hubbard1999,Kaspi2010a}.
The assumption is that in the dynamical regime expected to govern
the flow on the planet, the flow is accompanied by changes in the
density field, so that, given the gravity measurements, a static density
stratification together with a flow field could be found to best explain
the measurements.

To date, most models linking the wind (via the induced density anomalies)
to the gravity field to be measured by Juno, consider only flow that
is a projection of the observed cloud-level wind \citep[e.g.,][]{Hubbard1999,Kaspi2010a,Kaspi2013a,Zhang2015,Kaspi2016}.
Some assume full cylindrical flow while others allow for the wind
to decay with depth. However, none of the models included the possibility
of an internal flow that is decoupled from the surface-bound winds.
In addition, these models were able to calculate the gravitational
moments from a given flow field, but did not offer any methodology
for the inverse problem. In another study \citep{Galanti2016}, an
adjoint based inverse method was developed to relate the expected
gravity measurements to the flow underneath the cloud-level. It was
shown that given an measured gravity field the penetration depth of
the observed could-level wind could be recovered, even in cases where
this depth varies with latitude. The method also allows for measurement
errors to be incorporated, and uncertainties in the solution could
be calculated.

In this study, we explore the possibility of complex wind dynamics
that include both the surface bound wind, and a deep flow that is
completely detached from the flow above it. The methodology developed
in this study is a continuation of that presented in \citet{Galanti2016}.
There, the adjoint method was introduced and simple wind structures
were simulated and then shown to be invertible by the adjoint model
given the gravity moments. Here, we consider more complex flow cases,
and rigorously quantify the uncertainty in the adjoint solution and
the inevitability limits. The manuscript is organized as follows:
in section~\ref{sec:Methods} we describe the model and methods used
to calculate the complex flow structures, in section~\ref{sec:Results}
we discuss the various experiments performed, and conclusions are
given in section~\ref{sec:Conclusion}.

\section{Methods\label{sec:Methods}}

\subsection{The thermal wind model}

The dynamical model relating the flow on Jupiter to the density and
gravitational moments, is similar to the one used in \citet{Galanti2016}.
The model relates the flow field to the density field via the thermal
wind equation \citep{Kaspi2010a}. It assumes the dynamics to be in
the regime of small Rossby numbers, where the flow to leading order
is in geostrophic balance, therefore thermal wind balance holds
\begin{equation}
\left(2\Omega\cdot\nabla\right)\left[\widetilde{\rho}\mathbf{u}\right]=\nabla\rho'\times\mathbf{g_{0}},\label{eq: thermal wind}
\end{equation}
where $\Omega$ is the planetary rotation rate, $\widetilde{\rho}(r)$
is the background density field, $\mathbf{u}(\mathbf{r})$ is the
3D velocity, $\mathbf{g_{0}}\left(r\right)$ is the mean gravity vector
and $\rho'\left(r,\theta\right)$ is the dynamical density anomaly
\citep{Pedlosky1987,Kaspi2009}. The calculation takes advantage of
a known mean static density $\widetilde{\rho}(r)$ and gravity $\mathbf{g_{0}}\left(r\right)$
, calculated using the method of \citet{Hubbard1999}. In this study
we assume the flow is in the zonal direction only and does not vary
with longitude, so that $\mathbf{u}=u(r,\theta)\hat{e}_{\phi}$. The
model also assumes sphericity and excludes the effect of gravity anomalies
induced by the density anomalies. These specific assumptions were
shown to be a very good approximation of the full treatment of the
equations, which includes these effects \citep{Galanti2016b}. Moreover,
the thermal wind model was also shown to be in good agreement with
a more complete oblate potential theory model \citep{Kaspi2016}.

In the model used here, a modification was applied to the version
used in \citet{Galanti2016}. In a recent study, \citet{Kong2016}
showed that when asymmetry between the northern and southern hemisphere
winds exists, a more accurate solution is achieved when solving separately
for the two hemispheres. Following their conclusion, the numerical
derivative in latitude (\textit{lhs} of Eq.~\ref{eq: thermal wind})
is computed separately for the two hemispheres.

The dynamically induced gravitational moments are calculated using
the density solution $\rho$' from the thermal wind model, by integrating

\begin{equation}
J_{n}=-\frac{2\pi}{Ma^{n}}\intop_{0}^{a}r'^{n+2}dr'\intop_{-1}^{1}P_{n}\left(\mu'\right)\rho'\left(r',\mu'\right)d\mu',\label{eq: dynamical zonal harmonics}
\end{equation}
where $M$ is the mass of Jupiter, $a$ is the planet radius, $P_{n}$
are the Legendre polynomials, and $\mu=\cos\theta$. In the experiments
presented here we use the same model to generate both the 'observations',
denoted $J_{n}^{o}$, and the model solutions, denoted $J_{n}^{m}$.

\subsection{Construction of the surface-bound flow and deep flow }

For the upper surface bound flow (a flow that is manifested in the
cloud-level winds), we follow here the methodology of \citet{Galanti2016},
in which the observed cloud-level winds are projected along cylinders
parallel to the axis of rotation, and set to decay toward the high
pressure interior. The zonal wind field has the general form

\begin{equation}
U_{{\rm {surf}}}\left(r,\theta\right)=u_{0}\exp\left(\frac{r-a}{H(\theta)}\right),\label{eq:suface wind field}
\end{equation}
where $u_{0}\left(r,\theta\right)$ are the observed cloud-level zonal
winds extended constantly along the direction of the axis of rotation,
$a$ is the planet radius, and $H(\theta)$ is the latitudinal dependent
e-folding decay depth of the cloud level wind. The latitude dependent
$H$ is defined as a summation over Legendre polynomials

\begin{equation}
H\left(\theta\right)=\sum_{i=1}^{N_{H}}h_{i}P_{i-1}(\theta),\label{eq:depth of wind}
\end{equation}
where $P_{i}(\theta)$ are the Legendre polynomials, $h_{i}$ are
the coefficients by which the shape of $H(\theta)$ is determined,
and $N_{H}$ is the number of functions to be used. Such formulation
allows for a solution to be found separately for different spatial
scales of the winds and its resulting gravity signals. 

\begin{figure}[t!]
\begin{centering}
\includegraphics[width=0.65\paperwidth]{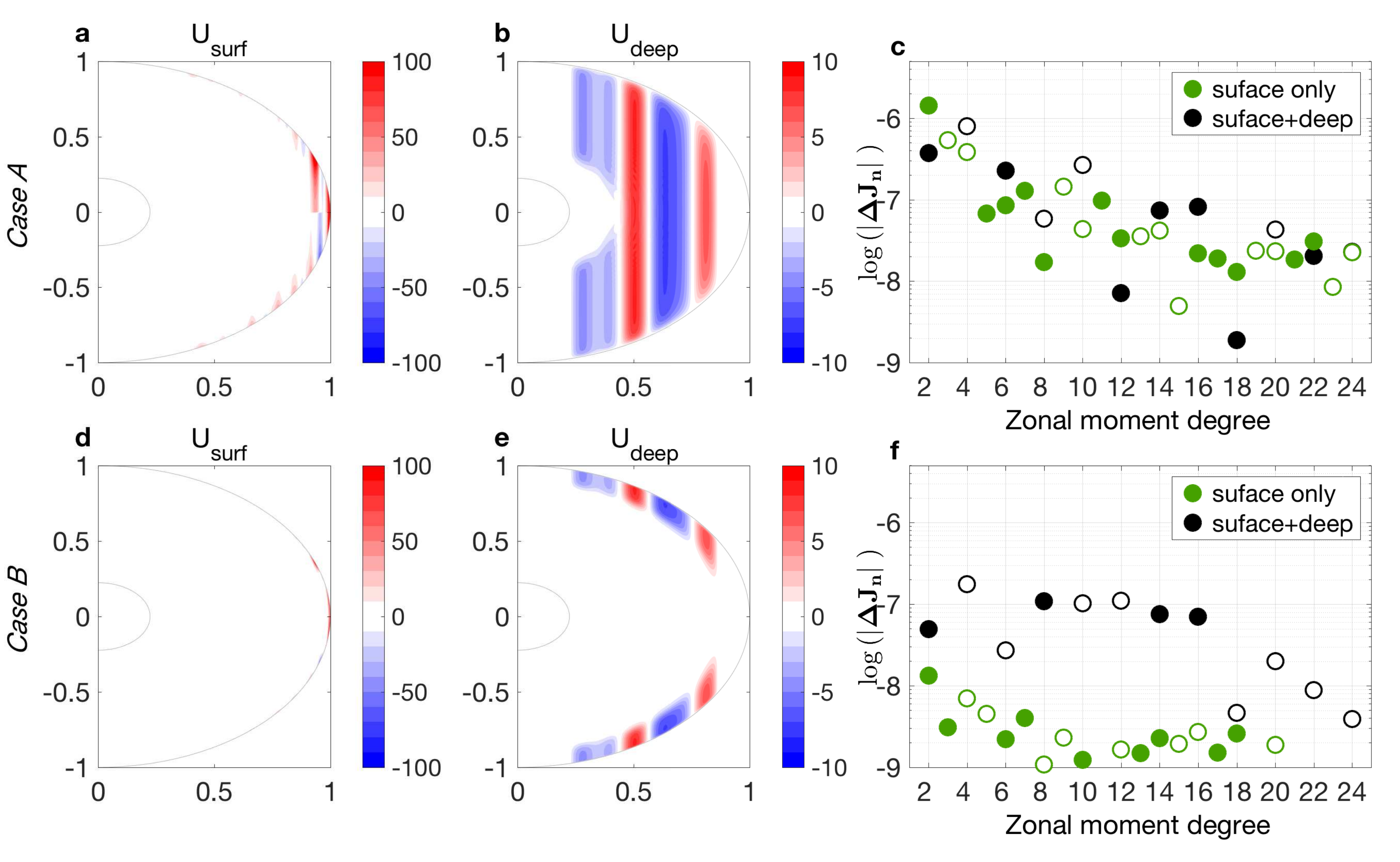}
\par\end{centering}
\caption{\label{fig: simulated wind and gravity}The simulated flow and resulting
gravitational moments for \textit{case A} (upper panels) and \textit{case
B} (lower panels). (a,d) Surface bound flow, (b,e) deep flow, and
(c,f) the gravitational moments resulting from surface flow only (green
dots), and combined surface and deep flow (black dots). }
\end{figure}

Next, we set a possible deep flow. The physical assumption taken is
that the observed flow pattern follows cylinders parallel to the planet
axis of rotation. This flow structure emerges in many studies in which
a general circulation model was forced by an internal heat source
\citep[e.g.,][]{heimpel2011,Christensen2002,Kaspi2009}. We also demand
that no deep flow exists inside the cylinder whose radius equals the
static core region assumed in the thermal wind model, i.e., when $l<\text{l}_{{\rm I}}$,
where $l=r\cos(\theta)$ is the distance from the axis of rotation,
and $l_{{\rm I}}=14,500$~km is the thermal wind model inner radius
(see Fig.~\ref{fig: simulated wind and gravity}b). Similarly, we
demand that no deep flow exits outside of $l_{{\rm O}}=60,000$~km
(equivalent to cutoff at latitude 30), the distance from axis of rotation
outside which the observed cloud level wind is dominating the interior
and no decoupling exists. This ensures that the deep flow does not
interfere with the strong surface jets in the equatorial region in
cases where they extend deep. The deep flow is set as
\begin{eqnarray}
U_{{\rm cyl}}(r,\theta) & = & \begin{cases}
l<l_{{\rm I}} & 0\\
l_{{\rm I}}<l<l_{{\rm O}} & \sum_{n=1}^{N_{U}}u_{n}\sin\left(\frac{n\pi(l-l_{{\rm I}})}{l_{{\rm O}}-l_{{\rm I}}}\right)\\
l>l_{{\rm O}} & 0
\end{cases},\label{eq:deep wind cylinders}
\end{eqnarray}
where $u_{1},...,u_{{\rm N_{U}}}$ are the magnitudes assigned to
the sinuous functions, and $N_{U}$ is the number of functions used.
This gives a flow structure that is function of $l$ only, and whose
value is zero at $l=l_{{\rm I}}$ and $l=l_{{\rm O}}$. 

Next, we demand that the deep flow decay toward the interior with
the function

\begin{eqnarray*}
D(r,\theta) & = & \frac{1}{2}\tanh\frac{r-r_{{\rm D}}}{\delta a_{{\rm D}}}+1,
\end{eqnarray*}
where $r_{{\rm D}}$ is the decay depth, and $\delta a_{{\rm D}}=2000$~km
is the decay scale. This enables the inclusion of a physical constraint
that the deep flow decays below a certain depth.

Finally, we demand that the deep flow is completely decoupled from
the surface-bound flow. For simplicity, we choose the decay function
to complement the decay function of the surface wind (Eq.~\ref{eq:suface wind field}),
so that the deep flow decays to zero at the planet surface. The total
deep flow is set as 

\begin{eqnarray}
U_{{\rm deep}}(r,\theta) & = & D\cdot\left[1-\exp\left(\frac{r-a}{H(\theta)}\right)\right]\cdot U_{{\rm cyl}}.\label{eq:deep wind field}
\end{eqnarray}

Using Eqs.~\ref{eq:suface wind field} and \ref{eq:deep wind field}
we set the total simulated wind field

\begin{eqnarray}
U & = & U_{{\rm surf}}(r,\theta)+U_{{\rm deep}}(r,\theta).\label{eq:total wind field}
\end{eqnarray}

\subsection{Simulated wind field and gravitational moments\label{subsec:Simulated-wind-field}}

Until the gravity observations from Juno will arrive, we can use the
thermal wind model to simulate the observed field \citep{Galanti2016}
given a surface-bound wind and a deep flow. The free parameters adjustable
when setting the total flow field are the depth of the surface wind
$H(\theta)$ based on the coefficients $h_{i}$, the structure of
the deep wind based on the coefficients $u_{i}$, and the depth of
the deep flow $r_{{\rm D}}$. Using these parameters, we define two
distinctly different scenarios, denoted as \textit{case A} and \textit{case
B}. These cases are chosen to illustrate potential observations made
by Juno.

In \textit{case A}, we set the surface wind depth coefficients to
$h_{1}=4000,\,h_{3}=-2000$~km (with all others set to zero). The
deep flow coefficients are set to $u_{1}=1,\,u_{4}=5,\,u_{5}=-3,\,u_{8}=2$
m~s$^{-1}$ (with all others set to zero). The depth of the deep
flow is limited to be outside of $a_{{\rm D}}=30,000$~km from center
of the planet. The resulting wind structure and the gravitational
moments calculated using the thermal wind model are shown in Fig.~\ref{fig: simulated wind and gravity}a,b,c.
The surface bound winds (Fig.~\ref{fig: simulated wind and gravity}a)
are pronounced mostly in the equatorial region where they penetrate
all the way to the equatorial plane. Its effect on the gravity moments
(Fig.~\ref{fig: simulated wind and gravity}c, green dots) is substantial.
The deep flow (Fig.~\ref{fig: simulated wind and gravity}b) has
a structure of positive zonal velocity in the low latitudes, a width
negative flow in the mid-latitudes, then again a strong positive flow
in higher latitudes, and finally a weak negative jet in the high latitudes.
Note that the strength of the deep flow is set to be an order of magnitude
smaller than the surface winds. The effect of the deep flow on the
gravitational moments is clear (Fig.~\ref{fig: simulated wind and gravity}c,
black dots) - some the even moment values are increased (for example,
$J_{4}$), while some values are decreased (for example, $J_{2}$),
and in some cases the even the sign is changed (for example, $J_{8}$).
Note that the odd moments ($J_{3,5,..})$ are not being modified by
the deep flow, since it is symmetric between the northern and southern
hemispheres. 

In \textit{case }B, we set the surface wind depth coefficients to
be 10 times smaller than in \textit{case A}. The result (Fig.~\ref{fig: simulated wind and gravity}d)
is that the surface wind is strongly limited to the surface, and its
affect on the gravitational moments is very small (Fig.~\ref{fig: simulated wind and gravity}f,
green dots). The deep flow coefficients are set as in \textit{case
A}, but the depth of the deep flow is now limited to $a_{{\rm D}}=0.9a\eqsim63,000$~km
(Fig.~\ref{fig: simulated wind and gravity}e), thus confined to
a much narrower region. The gravitational moments resulting from the
deep flow (Fig.~\ref{fig: simulated wind and gravity}f, black dots)
are smaller than in \textit{case }A\textit{, }but relative to the
surface wind, the deep flow is now dominating the gravity field. We
will use the total gravitational moments (black dots in Fig.~\ref{fig: simulated wind and gravity}c
for \textit{case A}, and those in Fig.~\ref{fig: simulated wind and gravity}f
for \textit{case B}) to simulate the observed field to be measured
by Juno, and denote them $J_{{\rm n}}^{o}$.

\subsection{Control variables and cost function definition }

The control variables we aim to optimize are the parameters defining
the depth of the surface wind $h_{1},...,h_{{\rm N_{H}}}$, the parameters
defining the structure of the deep flow $u_{1},...,u_{{\rm N_{U}}}$,
and the depth of the deep flow $a_{{\rm D}}$. Since each variable
has different units, the problem is best conditioned when the total
control vector is composed from the different parameters normalized
by their typical values. We define the control vector as 

\begin{eqnarray*}
\overrightarrow{\mathbf{X}_{{\rm C}}} & = & \left\{ \left[h_{1},...,h_{{\rm N_{H}}}\right]/h_{{\rm nor}},\left[u_{1},...,u_{{\rm N_{U}}}\right]/u,a_{{\rm D}}/a_{{\rm nor}}\right\} ,
\end{eqnarray*}
where $h_{{\rm nor}}=10^{7}$~m, $u_{{\rm nor}}=10$~m~s$^{-2}$,
and $a_{{\rm nor}}=10^{8}$~m. In the optimization procedure, the
values of the normalized control variables are limited to the range
of $-1$ to $1$, aside from the value for $a_{{\rm D}}/a_{{\rm nor}}$
is that limited between $0$ and $1$.

The cost function is defined similarly to \citet{Galanti2016}, as
the weighted difference between the model calculated moments and those
measured, with an additional penalty term that ensure that initial
guess does not affect the solution
\begin{eqnarray}
L & = & \mathbf{{\displaystyle \mathbf{\left(\mathbf{J^{m}}-\mathbf{J}^{o}\right)^{T}}W\left(\mathbf{J^{m}}-\mathbf{J}^{o}\right)}}+\epsilon\mathbf{X_{C}^{\mathbf{\mathbf{T}}}X_{\mathbf{\mathbf{C}}}},\label{eq:cost_function}
\end{eqnarray}
where $\mathbf{J}^{\mathbf{\mathbf{m}}}$ is the $N$ size calculated
model solution , $\mathbf{J}^{\mathbf{o}}$ is the observed one, and
$\mathbf{W}$ is a diagonal matrix of size $N\times N$ with weights
given to each moment $W_{ii}=4\times10^{16}$, representing simulated
uncertainties of $5\times10^{-9}$. The second term in Eq.~\ref{eq:cost_function}
act as a penalty term whose purpose is to ensure that the optimized
solution is not affected by the initial guess, or any part of the
control vector that do not affect the difference between the calculated
and observed gravity moments. An extensive discussion of this issue
(also known as the null space of the solution) can be found in \citet{Galanti2016b}.
The value of the parameter $\epsilon$ is set according to the initial
value of the cost function, so it affects the solution only when the
cost function is reduced considerably. The form of the penalty term
is set to penalize any non-zero value of the control variable $\mathbf{X_{C}}$
since we have no prior knowledge of either the depth of the surface
bound wind.

\subsection{Analysis of uncertainties in model solution\label{subsec:Analysis-of-uncertainties}}

When estimating a solution for the gravity field that best matches
the simulated (eventually, the observed) one, it is important to estimate
the uncertainties associated with the solution. These uncertainties
arise because the observations have uncertainties associated with
them, and therefore the combined range of the observation uncertainties
lead to uncertainties in the optimized variables. The control variable
uncertainties are derived from the the Hessian matrix $\mathbf{G}$
(second derivative of the cost function $L$ with respect to the control
vector $\mathbf{X}_{{\rm C}}$, see \citealt{Galanti2016}). Inverting
the Hessian matrix $\mathbf{G}$, we get the error covariance matrix
$\mathbf{\mathbf{C}}$. This matrix includes the error covariance
associated with combination of each two control variables (off diagonal
terms), and the variance of each one (diagonal terms). Physically,
the covariance matrix indicates to the formal uncertainties in the
control variables given the uncertainties of the observations (weights
$\mathbf{W}$ in the cost function). The larger the uncertainties
in the observations are, the smaller are the weights in the cost function,
and the larger the uncertainties in the control variables.

This information, however, does not give a direct estimate of the
physical parameters we are interested in. For example, the depth of
the surface wind (Eq.~\ref{eq:depth of wind}) is expressed in our
calculations as a summation over Legendre polynomials whose coefficients
are the control variables, and the information from the error covariance
matrix $\mathbf{C}$ is about them. However, our interest is in the
uncertainties associated with the depth of the surface wind as function
of latitude so that the information from the error covariance matrix
needs to be converted into information about the depth of the wind.
Consider a case where no correlation exists between errors of one
control parameter to another, so that the covariance matrix $\mathbf{C}$
has non zero values only on the diagonal, representing the variance
of each control variable error. In such a case, one can generate many
realizations of the errors, based on a normal distribution with standard
deviation taken from the diagonal terms of $\mathbf{C}$. Using Eq.~\ref{eq:depth of wind}
these realizations can be converted to the depth of the wind at each
latitude, so that for each latitude a distribution of errors is obtained.
From that distribution an estimation of the error could be calculated,
for example based on the $1^{th}$ standard deviation. However, in
all experiments discussed in this study the off diagonal terms in
$\mathbf{C}$ are substantial and cannot be neglected. Therefore,
a method should be derived for how to generate realizations of the
control variables errors, so that their covariance will satisfy $\mathbf{C}$.

By definition, the error covariance matrix has the form 

\begin{eqnarray}
\mathbf{C} & \equiv & \frac{\mathbf{X}\mathbf{X}^{T}}{N},\label{eq:covariance matrix}
\end{eqnarray}
where $\mathbf{X}=\left\{ \overrightarrow{\mathbf{X}_{1}},...,\overrightarrow{\mathbf{X}_{\mathbf{N}}}\right\} $
is the matrix whose rows contain $N$ realizations of the control
variables errors. Eigen decomposing of $C$ gives

\begin{eqnarray}
\mathbf{C} & = & \mathbf{S\Lambda}\mathbf{S}^{T},\label{eq:eigenC}
\end{eqnarray}
where $S$ is the matrix composed from the eigenvectors, and $\Lambda$
is a matrix with the eigenvalues on its diagonal. Note that $S^{-1}=S^{T}$
since $C$ is positive definite. Using Eq.~\ref{eq:covariance matrix}
we get

\begin{eqnarray}
\mathbf{S}^{T}\mathbf{X}\mathbf{X}^{T}\mathbf{S} & = & \mathbf{\Lambda}N.\label{eq:eigC1}
\end{eqnarray}
Defining a new set of error realizations $\mathbf{Y}=\left\{ \overrightarrow{\mathbf{Y}_{1}},...,\overrightarrow{\mathbf{Y}_{N}}\right\} $
, such that $\mathbf{Y}=\mathbf{S}^{T}\mathbf{X}$, we get

\begin{eqnarray}
\mathbf{Y}\mathbf{Y}^{T} & = & \mathbf{\Lambda}N,\label{eq:eigC2}
\end{eqnarray}
 so that the new control variable errors are uncorrelated. 

We can now generate $N$ realizations, where each error is based on
normal distribution with a standard deviation taken from $\Lambda$.
Each realization $\overrightarrow{\mathbf{Y}_{i}}$ is then converted
back to $\overrightarrow{\mathbf{X}_{i}}$ using $\overrightarrow{\mathbf{X}_{i}}=\mathbf{S}\overrightarrow{\mathbf{Y}_{i}}$.
The number of realizations $N$ can be determined from the requirement
that the realizations $\mathbf{X}$ satisfy Eq.~\ref{eq:covariance matrix}
to a certain degree. Requiring that the difference between the first
singular value of $C$ and the first singular value of $\mathbf{X}\mathbf{X}^{T}/N$
is less than one percent, we find that $N$ should be at least $500$.
Note that an equivalent to this method would be to use a Monte-Carlo
simulation to generate $\mathbf{X}$ directly based on $\mathbf{C}$,
but that would be computationally less efficient.

These realizations $\mathbf{X}$, satisfying the full error covariance
matrix \textbf{$\mathbf{C}$}, are converted to the actual physical
variables using Eq.~\ref{eq:depth of wind} for the depth of the
surface wind

\begin{equation}
H_{i}^{{\rm err}}\left(\theta\right)=\sum_{j=1}^{N_{H}}h_{{\rm nor}}X_{i}(j)P_{j-1}(\theta),\label{eq:depth of wind-std}
\end{equation}
and Eqs.\ \ref{eq:deep wind cylinders} and \ref{eq:deep wind field}
for the deep flow structure

\begin{eqnarray}
U_{i}^{{\rm err}}(r,\theta) & = & \begin{cases}
l<l_{{\rm I}} & 0\\
l_{{\rm I}}<l<l_{{\rm O}} & D\cdot\left[1-\exp\left(\frac{r-a}{H(\theta)}\right)\right]\\
 & \cdot\sum_{n=1}^{N_{U}}u_{{\rm nor}}X_{i}(n+N_{H})\\
 & \sin\left(\frac{n\pi(l-l_{{\rm I}})}{l_{{\rm O}}-l_{{\rm I}}}\right)\\
l>l_{{\rm O}} & 0
\end{cases}.\label{eq:deep wind-std}
\end{eqnarray}
Finally, the standard deviation (uncertainties) are calculated for
the physical variables $H_{i}^{{\rm err}}\left(\theta\right)$ and
$U_{i}^{{\rm err}}(r,\theta)$ from their $N$ realizations. Note
that the uncertainties of $H(\theta)$ are a function of latitude,
and those of $U(r,\theta)$ are a function of both latitude and depth.

\section{Results\label{sec:Results}}

We examine here the two simulated flow structures, \textit{case�A}
and \textit{case B}, under two distinctly different physical assumptions
. First, we use the same model for generating the simulated moments
and for finding the flow structure (section~\ref{subsec:Optimization-under-the}).
There we analyze our ability to reach a solution and the uncertainties
associated with it, for several combinations of control parameters.
Second, we look for a solution with a modified model in which the
physical constraints on the deep flow are completely relaxed (section~\ref{subsec:Optimization-with-unconstrained}).
This experiment serves as an end point to our ability to invert the
gravity moments into a flow field.

\subsection{Optimization under the same physical assumptions\label{subsec:Optimization-under-the}}

\subsubsection{\textit{Case A -} extended deep flow and deep surface wind  \label{subsec:Deep-flow-deep-surface}}

\begin{figure}[t!]
\begin{centering}
\includegraphics[scale=0.27]{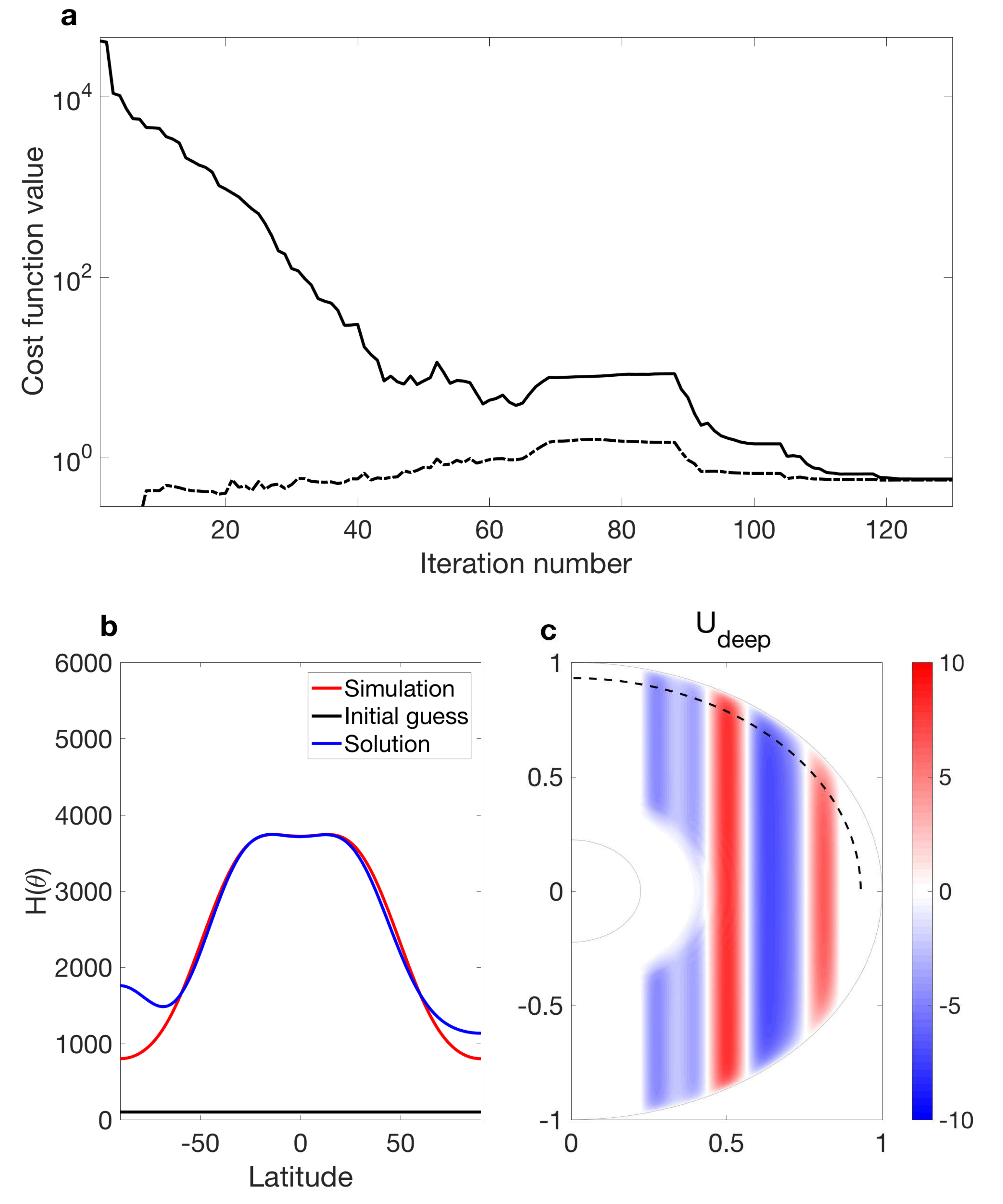}
\par\end{centering}
\caption{\label{fig: optimization - case A} Model optimization for \textit{case
A}. (a) The reduction of the cost function (solid) and its contribution
from the second term (dash-dotted), (b) Simulated, initial guess and
solution for the depth of the surface wind (red, black, and blue,
respectively), and (c) the solution for the deep flow (dashed line
shows the section analyzed in Figs.~\ref{fig: errors case A}b,d
and \ref{fig: errors case B}b,d).}
\end{figure}

We start by optimizing the model solution, compared to the one simulated
in \textit{case A}, using $N_{{\rm H}}=10$ functions for the depth
of the surface wind, $N_{{\rm U}}=10$ functions for the structure
of the deep wind, and set the depth of the deep winds $a_{{\rm D}}$
to be fixed. The total number of control variables is therefore 20.
As initial guess, we set $h_{1}=100$~km and all other control variables
to zero, so that initial guess gravity field resulting form the very
shallow surface wind and no deep wind is extremely small compared
to the simulated one. The results of the optimization are shown in
Fig.~\ref{fig: optimization - case A}. The reduction in the cost
function value (Fig.~\ref{fig: optimization - case A}a) indicates
to the different stages of the optimization. In the first stage (iterations
1 to 40), the reduction is mostly due to the adjustment of the lower
coefficients of the deep flow and depth of surface wind. Then, a rearrangement
is done in which the values of the higher coefficients are getting
much higher values but with little affect on the cost function (iterations
40 to 90), and the higher modes are adjusted close to the simulated
values. The depth of the surface wind, shown in Fig.~\ref{fig: optimization - case A}b,
is optimized from the initial guess (black line) to the model solution
(blue line), which is very close to the simulated depth (red line).
The deep flow structure (Fig.~\ref{fig: optimization - case A}c)
is almost identical to the simulated flow (Fig.~\ref{fig: simulated wind and gravity}b).
The small differences between the model solution and the simulation
are due to the setup of the termination conditions in the optimization
procedure, and points to the flatness of the cost function in the
vicinity of the global minimum. Note that the optimization is not
sensitive to the choice of the control variables initial guess (not
shown).

\begin{figure}[t!]
\begin{centering}
\includegraphics[scale=0.27]{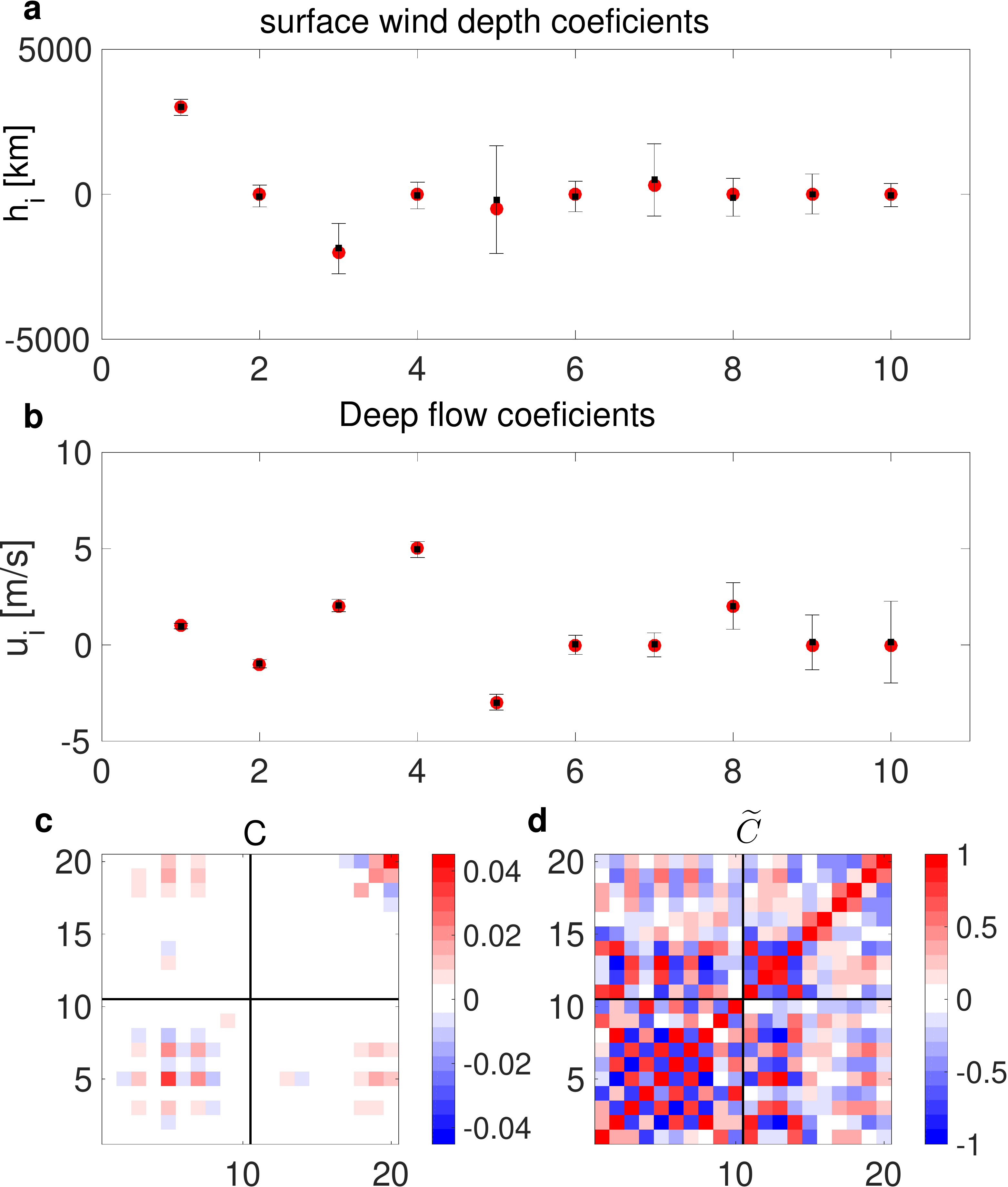}
\par\end{centering}
\caption{\label{fig: covariance - case A} Model solution (black dots) and
error standard deviation (error bars) for (a) The coefficients of
the surface wind depth, and (b) the coefficients of the deep flow
structure. Red dots are the simulation. (c) The error covariance matrix
where entries 1-10 are for the coefficients defining the depth of
the surface wind, and entries 11-20 are for the coefficients defining
the structure of the deep flow. (d) The normalized error covariance
matrix showing the correlation between the variables.}
\end{figure}

Next, we examine the solution for the control variables and the uncertainty
associated with them. Similar to the solution presented in Fig.~\ref{fig: optimization - case A},
the solution for the surface wind depth coefficients (Fig.~\ref{fig: covariance - case A}a,
black dots) and deep flow coefficients (Fig.~\ref{fig: covariance - case A}b,
black dots) are very close to the those used in the simulation (red
dots). Using the error covariance matrix (Fig.~\ref{fig: covariance - case A}c)
we can calculate the standard deviation for each variable, taking
the square root of the diagonal terms and renormalizing each variable.
These uncertainties are shown as error bars in Fig.~\ref{fig: covariance - case A}a,b.
It is apparent that the uncertainties depend strongly on the variables,
with some coefficients having small values and other much larger values.
For example, the standard deviation of the errors associated with
$h_{1}$ is $\sim280$~km while that associated with $h_{5}$ is
$\sim2000$~km. The standard deviation of the errors associated with
$A_{1}$ is $\sim0.15$ m~s$^{-1}$ while that associated with $A_{10}$
is $\sim2$ m~s$^{-1}$. Furthermore, there are strong correlations
between the different variables (off diagonal terms in Fig.~\ref{fig: covariance - case A}c),
which need to be taken into account when estimating the actual uncertainty
of the model solution. To illustrate this the normalized error covariance
matrix

\begin{eqnarray*}
\tilde{C}_{i,j} & = & \frac{C_{i,j}}{\sqrt{C_{i,i}C_{j,j}}},
\end{eqnarray*}
 is shown in Fig.~\ref{fig: covariance - case A}d. This matrix shows
the correlation between the variables, so that the diagonal terms
(self correlations) have a value of one, and off diagonal terms are
the correlation between each two variables. It is clear that many
strong positive and negative correlations exists, mainly between the
coefficients defining the depth of the surface wind (indices 1-10),
but also between these coefficients and those defining the structure
of the deep flow (indices 11-20).

\begin{figure}[t!]
\begin{centering}
\includegraphics[scale=0.3]{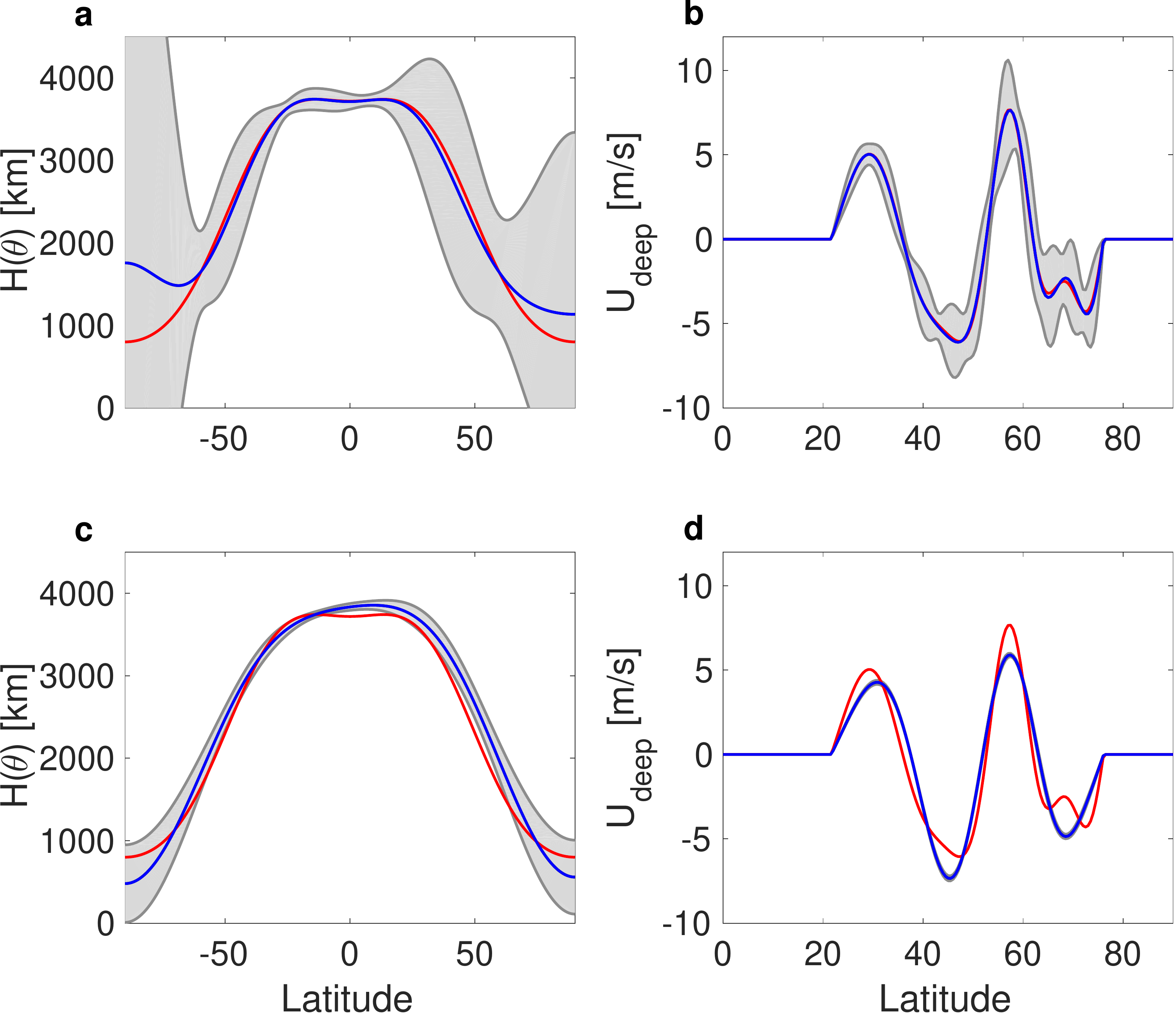}
\par\end{centering}
\caption{\label{fig: errors case A}(a) Model solution for the depth of the
surface wind (blue line), together with its uncertainties (shaded
area) and the simulated wind depth (red line). (b) Same as (a), but
for the deep flow structure along the radial distance of $0.93a$
(dashed line in Fig.~\ref{fig: optimization - case A}c). Panels
(c) and (d) are as (a) and (b), but for the case where the model is
optimized with $N_{{\rm H}}=5$ and $N_{{\rm U}}=5$.}
\end{figure}

Given that complexity, interpreting the error covariance matrix in
terms of the actual physical variables, the depth of the surface wind
$H(\theta)$ and the deep flow $U_{{\rm deep}}$, requires that all
information included in the covariance matrix is used. Following the
methodology presented in section \ref{subsec:Analysis-of-uncertainties},
we calculate the uncertainties associated with $H(\theta)$ and $U_{{\rm deep}}$.
In Fig.~\ref{fig: errors case A}a the standard deviation for the
errors in the depth of the surface wind is shown as shading on top
of the simulation. In the equatorial region, the model uncertainty
is confined to a few hundred kilometers but at high latitudes the
uncertainty rises to thousands of kilometers, implying that deviations
in the observed gravitational moments on the order of $5\times10^{-9}$
would results in the model inability to predict the depth of the surface
wind over the high latitudes. The model solution for the deep flow
structure along the radial distance of $0.93a\sim65,000$~km (dashed
line in Fig.~\ref{fig: optimization - case A}c) shows (Fig.~\ref{fig: errors case A}b)
that while the deep flow is well constrained in the outer part of
the planet (latitude $<30^{o}$), in the regions closer to the axis
of rotation the uncertainty increases, yet its' value is well below
the magnitude of the solution there. 

These results depend strongly on the number of optimized parameters;
the more parameters are used, the larger the uncertainty is \citep{Finocchiaro2010}.
To demonstrate this, consider a case where the model used for optimizing
the solution is based on a simpler structure of surface wind depths,
with $N_{{\rm H}}=5$, and a simpler deep wind structure, with $N_{{\rm U}}=5$
(Fig.~\ref{fig: errors case A}c,d). While the uncertainty in the
solution is now much smaller (shaded area), the solution itself (blue
lines) is less exact, especially in the equatorial region. This illustrates
the tradeoff between increasing the number of control variables (a
more exact solution), and the associated increased uncertainty. In
the specific case presented here, it is clear that for the deep flow
structure $N_{{\rm U}}=5$ does not provide enough spatial variability
since $A_{8}$ has a considerable contribution in the simulation (see
Fig.~\ref{fig: covariance - case A}b). The model solution (Fig.~\ref{fig: errors case A}d,
blue line) is missing a sizable part of the simulated flow structure
(red line). The depth of the surface wind, on the other hand, has
only little contribution from $h_{7}$, therefore the model solution
with $N_{{\rm H}}=5$ (Fig.~\ref{fig: errors case A}d, blue line)
is quite similar to the simulated one (red line).

Finally, we discuss briefly a couple of modified experiments. First,
a variant of the above experiment is to set the depth of the surface
bound wind to be 10 times smaller, thus making it more superficial.
Results show that it is more difficult to reconstruct the simulated
depth of the surface wind, but overall the results are similar, especially
for the deep flow. Uncertainties also are qualitatively similar. In
another variation, in addition to the surface depth and the structure
of the deep flow we also set as a control variable the depth for the
deep flow. The ability to reach the global minimum is this case is
degraded considerably and the solution depends on the initial guess.
In some cases a global minimum is reached, but in others the solution
is far from the simulation, not only in the depth of the deep flow,
but also in all other parameters. As discussed below, this is a less
of a problem in \textit{case B} where the depth of the deep flow is
restricted to relatively shallow levels. 

\subsubsection{\textit{Case B - }restricted deep flow and shallow surface wind\label{subsec:Deep-flow-shallow-surface}
 }

Next, we examine the characteristics of the optimization where the
simulated gravitational moments are based on \textit{case B,} where
the depth of the deep flow is much more limited, and where the surface
wind is shallow (section~\ref{subsec:Simulated-wind-field}, Fig.~\ref{fig: simulated wind and gravity}d,e,f). 

\begin{figure}[t!]
\begin{centering}
\includegraphics[scale=0.3]{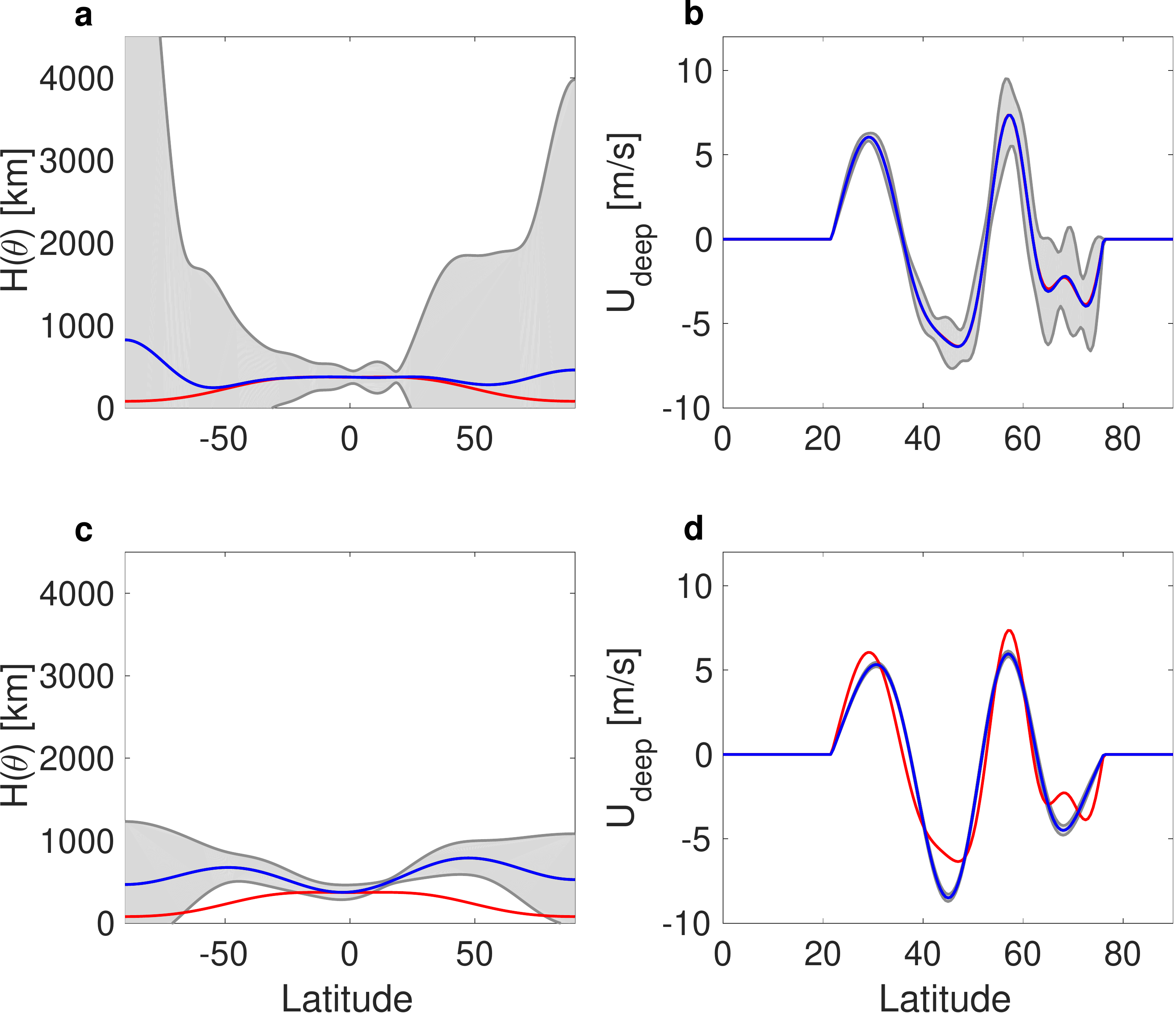}
\par\end{centering}
\caption{\label{fig: errors case B} Results for \textit{case }B -\textit{
}(a) Model solution for the depth of the surface wind (blue line),
together with its uncertainties (shaded area) and the simulated wind
depth (red line. (b) Same as (a), but for the deep flow structure
along the radial distance of $0.93a$ (dashed line in Fig.~\ref{fig: optimization - case A}c).
Panels (c) and (d) are as (a) and (b), but for the case where the
model is optimized with $N_{{\rm H}}=5$ and $N_{{\rm U}}=5$.}
\end{figure}

As in \textit{case A, }the control variables include the 10 surface
wind depth coefficients and the 10 deep flow coefficients. Later on,
we also consider the depth of the deep flow $a_{D}$ as a control
variable. Starting with an initial guesses similar to those used in
section \ref{subsec:Deep-flow-deep-surface}, the optimization is
able to reach a solution that is in general as good as the one achieved
for the experiment presented in section~\ref{subsec:Deep-flow-deep-surface}
(Fig.~\ref{fig: errors case B}). The solution reproduces well the
depth of the surface wind aside from the polar regions (panel a) and
the deep flow (panel b). Restricting the number of surface wind depth
coefficients and deep flow structure coefficients to 5 (panels c,d),
reduces the uncertainties but causes the solution to agree less with
the simulation. 

Several important differences from \textit{case A} arise. First, comparing
Fig.~\ref{fig: errors case A} and Fig.~\ref{fig: errors case B},
while the uncertainties for the deep flow are similar in both cases,
the uncertainties for the depth of the surface wind are larger in
\textit{case B}. This is true for both large or small number of coefficients
used (panels a and c). More importantly, since the depth of the surface
wind is now 10 times shallower, the even larger magnitude of the uncertainties
implies that aside from the equatorial region, it would be impossible
to place a lower limit on the depth (the gray zone reaches a depth
of zero), and the upper limit is now more than $1000$~km in most
latitudes (panel a). Even with the reduction of the number of coefficients
(panel c), the uncertainty is still much larger than the simulated
depth. 

On the other hand, including the optimization of the deep flow depth
$a_{{\rm D}}$ is now feasible in some cases. While in the equivalent
experiments discussed in section \ref{subsec:Deep-flow-deep-surface},
in inclusion of $a_{{\rm D}}$ resulted in a solution dependent on
the initial guess of the control variables, in \textit{case B }a global
solution very similar to the one shown in Fig.~\ref{fig: errors case B}
could be reached when the initial guess is $a_{{\rm D}}\gtrsim50,000$~km
(not shown). With that, setting the initial guess to lower values
results in the optimization reaching a local minimum that is far from
the simulated one.

\subsection{Optimization with unconstrained deep flow\label{subsec:Optimization-with-unconstrained}}

So far, the same model has been used for generating the simulation
and to search for the solution. We now examine a case where the model
used for finding the solution differs profoundly from the one used
to simulate the observations. An extreme test for the model ability
to reach a solution would be to relax the structure of the deep flow,
from cylinders to a general flow that has absolutely no restrictions
in both latitude and depth. Physically such a solution is likely unjustifiable,
but this serves as a good test for the model's ability to reach a
solution without any constraints. The implication to the adjoint model
and optimization process is profound. Now, in addition to the 10 control
variables of the depth of the surface wind, there are $N_{\theta}\times N_{r}$
control variables of $U_{{\rm deep}}$ (compared to the 10 control
variables used before to set the deep wind structure). While in the
above experiments we set $N_{\theta}=361$ and $N_{r}=174$ ($0.5$~deg
resolution in latitude and 10 vertical levels per scale height), here
a reduced resolution has to be employed, $N_{\theta}=91$ and $N_{r}=87$,
so that the total length of the control variable is $91\times87+10=7927$.
Even with such a reduced resolution, the numerical calculation of
the optimization (finding the search direction and step length at
each iteration, see \citet{Galanti2016} for details) requires considerable
computational resources. Note that since we are producing the 'observations'
and looking for the solution with the same model resolution, reducing
the resolution is self-consistent and does not affect the results.
However, when analyzing the Juno observations, numerical capability
to solve the high resolution version would have to be developed.

\begin{figure}[t!]
\begin{centering}
\includegraphics[scale=0.33]{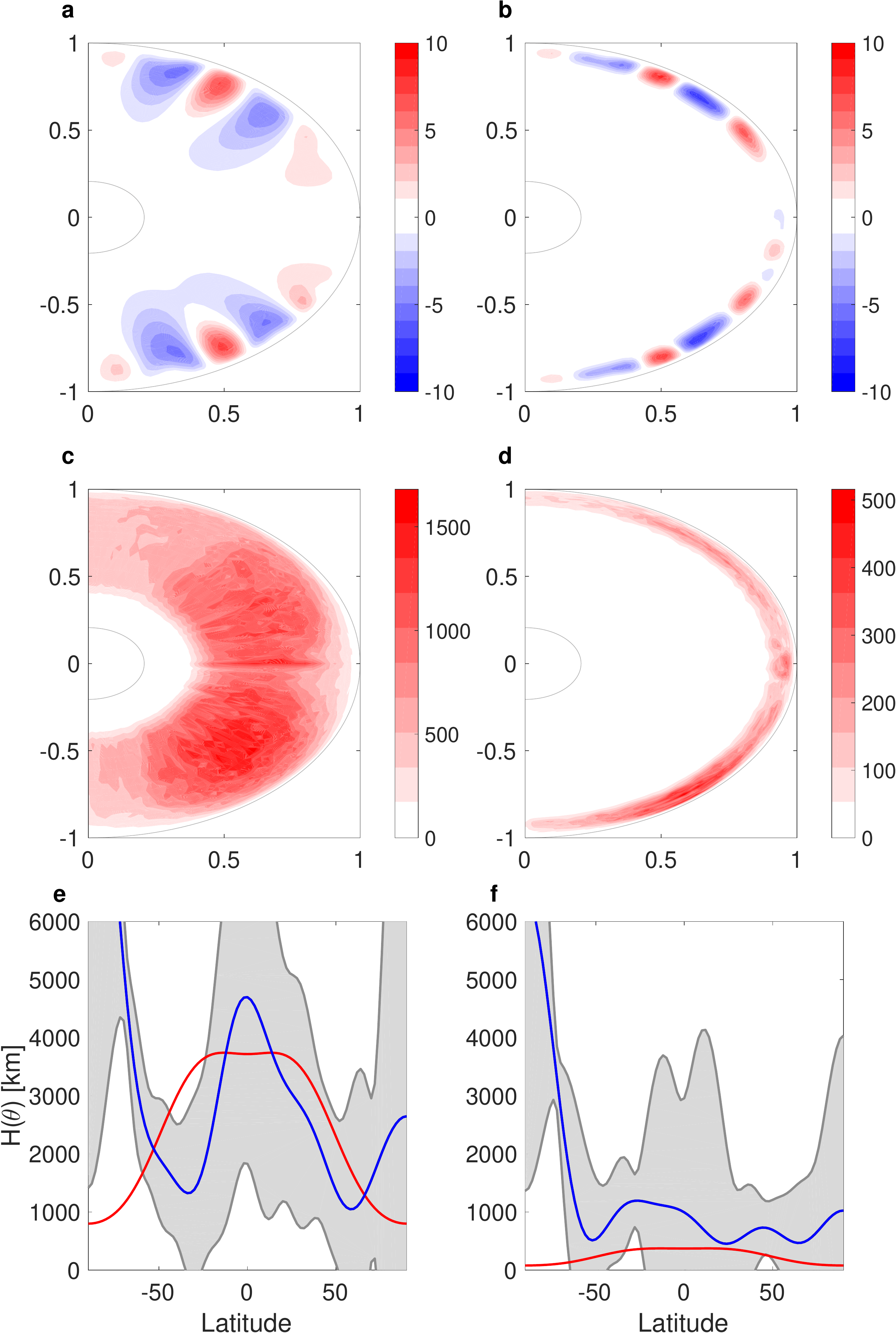}
\par\end{centering}
\caption{\label{fig: unconstrained model}Looking for an optimized solution
with the unconstrained deep flow, in both \textit{case A} (left panels),
and \textit{case B} (right panels). Shown are the solution for the
deep flows (a,b), the uncertainties associated with this solution
(c,d), and the solution for the depth of the surface wind (e,f) with
its uncertainties. }
\end{figure}

We consider here our ability to reach the solution in both simulations,
\textit{case A} and \textit{case B }(Fig.~\ref{fig: unconstrained model}).
A striking characteristic of the solution in both cases (Fig.~\ref{fig: unconstrained model}a,b)
is the structure of the deep flow is aligned mostly in the radial
direction. Now that the deep flow is no longer constrained to flow
parallel to the axis of rotation, it is the thermal wind balance and
the radial independence of the gravitational moments that set the
optimal flow. That said, it is encouraging that the overall structure
is similar to the simulated one (Fig.~\ref{fig: simulated wind and gravity}b,d).
Going from the poles to the equator, the pattern of negative, positive,
negative, and then positive flow, is apparent in both cases. The solution
in both cases also identifies the lack of deep flow close to the equator.
The depth of the deep flow, however, has different characteristics
in the two cases. In \textit{case A} (Fig.~\ref{fig: unconstrained model}a),
the deep flow is not extending all the way to a depth of $30,000$~km
as in the simulation, suggesting the the flow in that region (if existing)
could not be recovered in the model solution. This results form the
fact that the affect of the density on the gravitational moments is
decreasing with depth. In \textit{case B} (Fig.~\ref{fig: unconstrained model}b),
the solution extends in the entire region where the deep flow exists
in the simulation.

The major limitation however comes from the uncertainties (Fig.~\ref{fig: unconstrained model}b,d),
which extend over the entire region and have unphysical values of
more than $1,500\,{\rm m\,s^{-1}}$ in \textit{case A}, and more than
$500$ m~s$^{-1}$ in \textit{case B}. This is also the case for
the uncertainties in the solution for the depth of the surface wind
(Fig.~\ref{fig: unconstrained model}e,f), which is of the order
of several thousands of kilometers.

\section{Conclusion\label{sec:Conclusion}}

We develop a methodology to examine the upcoming high accuracy gravity
measurements by the Juno spacecraft. We allow the flow structure to
be as general as possible, with a deep flow which is completely decoupled
from a surface bound observed flow. The model is composed from a forward
dynamical model that relates the 3D flow to the density and gravity
fields, and an inverse model that, given the observed gravity field,
can trace back the complex flow.

In order to simulate possible observations of the gravity field, we
constructed two observational scenarios. In the first, the interior
flow is deep, and the cloud-level flow penetrates to a depth of $3000$~km.
Deeper flows in this case have a comparable effect on the gravitational
moments. In the second scenario, interior flow is confined to a relatively
narrow region, and the surface bound flow is shallow, penetrating
to a depth of only $300$~km. The gravitational moments in this case
are mostly affected by the deep flow, while the surface bound flow
has a negligible effect. These two cases were selected as representative
cases to allow examination of substantially different scenarios.

We examine the inverse model ability to reach a solution for both
scenarios. We start from an initial guess of no interior flow, and
close to zero depth of the surface bound flow. We then seek for a
solution that minimize the difference between the calculated and the
simulated gravity moments. For both simulations, \textit{case A }and
\textit{case B}, the optimized solution reproduces well the depth
of the surface wind (aside from the polar regions) and the deep flow
structure and amplitude. In both cases, optimizing when the number
of surface wind depth coefficients and deep flow structure coefficients
is restricted to 5, caused the uncertainties in the solution to be
reduced, but the solution agreed less with the simulation. Nonetheless,
several differences between the cases exist. The uncertainties for
the depth of the surface wind are larger in \textit{case B}. This
is true for both large or small number of coefficients used. More
importantly, since the depth of the surface wind in \textit{case B
}is 10 times shallower than \textit{case A}, even larger magnitude
of the uncertainties implies that aside from the equatorial region,
it would be impossible to place a lower limit on the depth. Overall,
in a case where the surface bound flow is on the order of $\sim100$~km,
and there exist a decoupled flow in the interior, it will be very
difficult to estimate the depth of surface flow. Even with the reduction
of the number of coefficients, the uncertainty is still much larger
than the simulated depth. 

Finally, we examined a case where the model used for finding the solution
differs considerably from the one used to simulate the observations.
We tested the model ability to reach a solution when the flow field
is free to have any possible form, so it is not restricted to cylindrical
shapes, and can have a general flow that has absolutely no restrictions
in both latitude and depth. Physically, it is hard to justify such
a solution, but this serves as a good test for the model's ability
to reach a solution without any constraints. In both simulations,
\textit{case A} and \textit{case B}, the structure of the solution
was aligned mostly in the radial direction and not parallel to the
axis of rotation as the simulated flow. Nevertheless, the overall
structure of the solutions was similar to the simulated one. The depth
of the deep flow did not match well the simulated one, and had different
characteristics in the two cases. A major limitation comes from the
uncertainties that extend over the entire region and have unphysical
values This was also the case for the uncertainties associated with
the solution for the depth of the surface wind. Their values was in
the order of several thousands of kilometers.

The novelty of the adjoint based inverse method presented here is
in the ability to identify complex flow dynamics given the expected
Juno measurements of gravity moments. Unlike any previous studies,
this model allows also for the existence of deep cylindrical flows
that have no manifestation in the observed cloud-level wind. Furthermore,
the flexibility of the adjoint method allows for a wide range of dynamical
setups, so that when new observations and physical understanding will
arise, these constraints could be easily implemented and used to better
decipher Jupiter flow dynamics.

~

\textit{Acknowledgments: }We thank Eli Tziperman and the members of
the Juno science team interiors working group for helpful discussions.
This research has been supported by the Israeli Ministry of Science
and the Minerva foundation with funding from the Federal German Ministry
of Education and Research. We also acknowledge support from the Helen
Kimmel Center for Planetary Science at the Weizmann Institute of Science.

 \bibliographystyle{elsarticle-harv} 

\bibliography{bibliography_main}

\end{document}